\documentstyle[preprint,prb,aps]{revtex}
\tighten
\begin{document}

\draft

\title{Odd Parity and Line Nodes in Heavy Fermion Superconductors}

\author{M.R. Norman}
\address{Materials Science Division, Argonne National Laboratory,
Argonne, IL 60439}

\date{\today}

\maketitle

\begin{abstract}

Group theory arguments have demonstrated that a general odd parity order
parameter cannot have line nodes in the presence of spin-orbit coupling.  In
this paper, it is shown that these arguments do not hold on the $k_z = \pi/c$
zone face of a hexagonal
close packed lattice.  In particular, three of the six odd parity
representations vanish identically on this face.  This has potential relevance
to the heavy fermion superconductor $UPt_3$.

\end{abstract}


\narrowtext

The symmetry of the order parameter of heavy fermion superconductors is still
an unresolved issue after over a decade's worth of work.\cite{review}  Even
the parity of the order parameter has not been determined.  At an early
stage, though, group theory arguments were given that limited the number of
possibilities for the order parameter.\cite{vg,blount}  In particular, Blount
showed that a general odd parity order parameter would not have line nodes
in the presence of spin-orbit coupling.\cite{blount}  Since experimental
evidence in many heavy fermion superconductors, especially $UPt_3$,  point to
the presence of line nodes in the order parameter, most theoretical models
assume an even parity order parameter.  In this paper, a review of this
argument is given and then a particular example is analyzed
where this argument fails.  This example is the $k_z=\pi/c$ zone face of a
hexagonal closed packed lattice, such as $UPt_3$, where it
will be shown that three of the six odd parity representations vanish.

In this paper, only the case of a hexagonal close packed lattice is treated.
The above argument of Blount is most easily illustrated by the use of basis
functions, a complete set of which were recently published by Yip and
Garg\cite{yip} (Blount's argument, though, is general and does not
depend on the use of basis functions).
For instance, consider the $E_{1u}$ ($\Gamma_5^-$)
representation.  Basis functions at the p-wave level are
$k_z ({\bf \hat{x}} \pm i{\bf \hat{y}})$ and $(k_x \pm ik_y){\bf \hat{z}}$
where ${\bf \hat{x}}$, ${\bf \hat{y}}$, and ${\bf \hat{z}}$ are basis functions
for $S$=1.  Although the first function does has a line of nodes,
in the presence of spin-orbit, the two functions will be mixed with one
another (since $S_z$ is no longer a good quantum number).  Therefore, in
general, only point nodes can occur.

Most odd parity models that have been discussed in connection with $UPt_3$
ignore this mixing effect (an example being the $E_{2u}$ model of
Norman\cite{norm1} and Sauls\cite{sauls} where only the $S_z$=0 component
is kept).  An exception was a recent model of Norman\cite{norm2} which treated
the strong spin-orbit limit of on-site pairing.  In this case, it was found
that the $\Gamma_6^-$ ($E_{2u}$) pair state vanished on the $k_z=\pi/c$ zone
face.  The question is whether this result is specific or can be generalized.

The first question to address is how Blount's argument was circumvented in this
case.  To do this requires an analysis of single particle wavefunctions.
The f electron part of the single particle wavefunctions is a linear
combination of $J$=5/2 functions (where $J$ denotes the total angular
momentum\cite{refj}).  In the case of a hexagonal close packed
lattice, there are two f atoms per primitive cell (separated by a
non-primitive translation vector).  Therefore, the wavefunction is of the
form $a^{n \vec{k}}_{\mu i} |\mu>_i$
where $n$ is a band index, $\mu$ a basis function (-5/2,-3/2,-1/2,1/2,3/2,5/2),
and $i$ a site index (1,2).
Consider the two symmetry planes $k_z=0$ and $k_z=\pi/c$ (these are the only
symmetry planes perpendicular to the c axis, and are of interest for
$UPt_3$ since line nodes perpendicular to c have been inferred experimentally).
For a particular
site, $i$, only functions differing by two units of angular momentum can mix.
This occurs since these planes are mirror planes relative to the operation
z $\rightarrow$ -z ($\sigma_h$) and the functions -5/2,-1/2,3/2 transform as -i
and -3/2,1/2,5/2 as +i under this operation.\cite{ah}  Thus, for a particular
site, the coefficients of either -5/2,-1/2,3/2 or -3/2,1/2,5/2 vanish.
For the $k_z=0$ case, the $a_{\mu}$ coefficients
which are zero on one site are also zero on the other site,
but for the $k_z=\pi/c$ case, they are ``staggered'' (that is, if -5/2,-1/2,3/2
are zero on site 1, then -3/2,1/2,5/2 will be zero on site 2). This
difference occurs since the factor $e^{i\vec{k}\cdot\vec{r}}$ will introduce
a relative phase between the two sites of 1 for $k_z=0$ and -1 for $k_z=\pi/c$.

Now consider basis functions for pairs of f electrons.  Since a center of mass
momentum of zero is assumed (that is, $\vec{k}$ and $-\vec{k}$ are paired),
these pairs are from representations at the
$\Gamma$ point of the zone ($\Gamma_1$ through $\Gamma_6$).  In addition,
the correct combination of these pair states corresponding to either even or
odd parity must be constructed.  This was first considered in the presence
of spin-orbit by Anderson.\cite{and}  At a general $\vec{k}$, there are two
Kramers degenerate states labeled $k$ and $PTk$ where $P$ is the parity
operator and $T$ the time reversal operator, corresponding to up spin and down
spin in the spin only case.  The analogous states at $-\vec{k}$
can be labeled as $Pk$ and $Tk$.  The even parity combination is then
$k,Tk-PTk,Pk$ and is a pseudo-spin singlet (corresponding to $S=0$ in
the spin only case).  For odd parity, there are three pseudo-spin combinations:
$k,Pk$ and $PTk,Tk$ and $k,Tk+PTk,Pk$ (corresponding to $S=1$ in the spin only
case).  These are conveniently relabeled as a vector, $\vec{d}$, with the
above three states corresponding to $-d_x+id_y$, $d_x+id_y$, and $d_z$.
Finally, to consider the full effect of the space group on the order
parameter, it is necessary to analyze the pair wavefunction in real space.
The cases of electons on the same site,
electrons separated by a non-primitive lattice vector, and electrons
separated by a primitive lattice vector have been treated by Appel and
Hertel\cite{ah} (it is from this work that the arguments below will be
obtained).  By construction, then, if something is proved for these three
cases, then it is true for the general pair wavefunction since all f atom sites
are connected by either a primitive or non-primitive lattice vector.

For on-site pairs, three of the six possible odd parity representations,
labeled even z representations
($\Gamma_1^-$, $\Gamma_2^-$, $\Gamma_5^-$), involve states with even $M_J$,
the other three, labeled odd z representations ($\Gamma_3^-$, $\Gamma_4^-$,
$\Gamma_6^-$), involve states with odd $M_J$.\cite{refm}  This along with the
statements in the above two paragraphs allows
one to trivially conclude the following (remembering that the operator $P$
interchanges sites 1 and 2 in the unit cell, whereas the operator $T$
interchanges $\mu$ and $-\mu$).  For $k_z=0$, using the relations satisfied by
the single particle wavefunctions, combinations like $k,Pk$ are
non-vanishing only for odd $M_J$ states whereas combinations like $k,Tk$
are non-vanishing only for even $M_J$ states.\cite{refpt}
Thus, for even z representations,
$d_x$ and $d_y$ vanish, whereas for odd z representations, $d_z$ vanishes.
These are the arguments appropriate to basis functions as discussed in the
first part of this paper since such basis functions involve expansions about
$\vec{k}=0$.  The situation, though, changes for the $k_z=\pi/c$ case
given the ``staggering'' of the single particle wavefunctions discussed above.
In this case, only even $M_J$ states are non-vanishing for both $k,Pk$
and $k,Tk$ combinations.  Therefore, for even z representations, all three
pseudo-spin components are in general non-zero on this face, whereas for odd z
representations, all three pseudo-spin components vanish identically.

The case of next near neighbor pairs (electrons separated by a primitive
lattice vector in the basal plane) turns out to be identical to the on-site
case.  This is expected, since a primitive lattice vector is involved.
Formally, the group for two fixed sites separated by a lattice vector
is $C_s$ composed of the identity, $E$, and $\sigma_h$.  There are two
representations of this group, $\Gamma_1$ (even z) and $\Gamma_2$ (odd z).
The former is only composed of even $M_J$ pairs, the latter of odd $M_J$
pairs.  When the full space group is considered (that is, all rotations of
the two sites plus their interchange), then $\Gamma_1$ leads to the even z
representations, $\Gamma_2$ to the odd z representations, and thus the
arguments of the above paragraph follow immediately.

The case of near neighbor pairs (electrons separated by a non-primitive
lattice vector) is somewhat different.  In this case, one electron is at
site 1 in the primitive cell, the other at site 2.  Again using the properties
of the single particle wavefunctions discussed above, for the $k_z=0$
case, only odd (even) $M_J$ states are non-vanishing for the combination $k,Pk$
($k,Tk$) just as before.  On the other hand, for the $k_z=\pi/c$ case,
only odd $M_J$ states are non-vanishing for both combinations (previously,
it was even $M_J$).  To complete the argument, though, one needs to know how
even z and odd z representations transform.  A key difference from before is
due to the c axis being a screw axis.  Thus, the operation
$\sigma_h$ must be followed by a non-primitive translation.  In the previous
paragraphs, this resulted in a phase factor of unity since the pairs $k,-k$
involved electrons either
both at atom site 1 or both at atom site 2 (modulo a primitive lattice vector).
In the present case, though,
one of the electrons is at site 1, the other at site 2, resulting in an
overall phase factor of $e^{ik_zc}$ where c is the lattice constant along the
c axis.  Thus the effect of $\sigma_h$ on a pair is $(-1)^{M_J}
e^{ik_zc}$.  For odd (even) z representations, then, $M_J$ must be odd (even)
for $k_z=0$ and even (odd) for $k_z=\pi/c$ to be non-vanishing.  Combining
this with the constraints of the single particle wavefunctions mentioned
earlier, one then
finds the same results as before.  That is, for odd z representations,
$d_x$ and $d_y$ are non-vanishing and $d_z$ vanishing for $k_z=0$, but
for $k_z=\pi/c$, all vanish; whereas for even
z representations, $d_x$ and $d_y$ are vanishing and $d_z$ non-vanishing
for $k_z=0$, but all are non-vanishing for $k_z=\pi/c$.

The above arguments have implications, at least for the case of $UPt_3$.
The calculated Fermi surface for $UPt_3$ has two of the five Fermi surface
sheets centered about the $k_z=\pi/c$ zone face (contributing about 43\%
to the total density of states)\cite{nabc} and both sheets are in good
agreement with deHaas-vanAlphen data.\cite{dhva}  Therefore, the existence
of line nodes in $UPt_3$ cannot be used as a criterion to differentiate
between even and odd parity pairing, since given the above arguments,
two of the five sheets will have line nodes for three of the possible six
odd parity representations, even if the pair state involves all three
components
of the d vector.

\acknowledgments

This work was supported by the U.S. Dept. of Energy,
Basic Energy Sciences, under Contract No. W-31-109-ENG-38.  The author
also acknowledges the support of the Aspen Center for Physics where this
work was begun.

\vfill\eject


\begin{references}
\bibitem{refj} The arguments of this paper do not depend on the value of $J$
or whether one or several $J$ are involved.
\bibitem{refm} $M_J$ is equal to $\mu+\mu^{\prime}$ where $\mu$ comes from the
expansion of $k$ and $\mu^{\prime}$ from the expansion of -$k$.  In general,
the
pair wavefunction will contain several $M_J$ as well as several $J$ (where $J$,
the total angular momentum, now refers to the pair), but for the two $k_z$
values considered here, odd and even $M_J$ are not mixed.
\end{references}
\end{document}